\documentclass[a4paper,twoside]{article}

\usepackage{verbatim}
\usepackage{multicol}
\usepackage{multirow}
\usepackage{pslatex}
\usepackage{apalike}
\usepackage{csquotes}
\usepackage[hidelinks]{hyperref} 
\usepackage{array}
\usepackage{footnote}
\newcolumntype{C}[1]{>{\centering\let\newline\\\arraybackslash\vspace{0.5pt}}m{#1}}

\usepackage{epsfig}
\usepackage{subcaption}
\usepackage{calc}
\usepackage{amssymb}
\usepackage{amstext}
\usepackage{amsmath}
\usepackage{amsthm}
\usepackage{SCITEPRESS}     

\begin{document}

\title{Evasive Windows Malware: Impact on Antiviruses and Possible Countermeasures}

\author{\authorname{C\'edric Herzog\sup{1}, Val\'erie Viet Triem Tong\sup{1}, Pierre Wilke\sup{1}, Arnaud Van Straaten\sup{1}, and Jean-Louis Lanet\sup{1}\orcidAuthor{0000-0002-4751-3941}}
\affiliation{\sup{1}Inria, CentraleSupélec, Univ Rennes, CNRS, IRISA, Rennes, France}
\email{\{f\_author, s\_author\}@inria.fr}
}


\abstract{The perpetual opposition between antiviruses and malware leads both parties to evolve continuously. On the one hand, antiviruses put in place solutions that are more and more sophisticated and propose more complex detection techniques in addition to the classic signature analysis. This sophistication leads antiviruses to leave more traces of their presence on the machine they protect. To remain undetected as long as possible, malware can avoid executing within such environments by hunting down the modifications left by the antiviruses. This paper aims at determining the possibilities for malware to detect the antiviruses and then evaluating the efficiency of these techniques on a panel of antiviruses that are the most used nowadays. We then collect samples showing this kind of behavior and propose to evaluate a countermeasure that creates false artifacts, thus forcing malware to evade.
}

\onecolumn \maketitle \normalsize \setcounter{footnote}{0} \vfill


\label{sec:introduction}

\section{\uppercase{Introduction}}

\noindent There is a permanent confrontation between malware and antiviruses (AVs).\footnote{\href{https://www.av-test.org/en/statistics/malware/}{https://www.av-test.org/en/statistics/malware/}}
On the one hand, malware mainly intend to infect devices in a short amount of time, while remaining undetected for as long as possible. On the other hand, of course, AVs aim at detecting malware in the fastest way possible. 

When a new malware is detected, AVs lead further analysis to produce a signature and keep their database up-to-date quickly. Malware authors, willing to create long-lasting malware at low cost, can then use simple methods to avoid being detected and analyzed deeply by these AVs. To this end, a malware can search for the presence of traces or artifacts left by an AV, and then decide whether or not they execute their malicious payload. We call such a malware an \emph{evasive malware}.

This article focuses on the evaluation of the evasion techniques used by Windows malware in the wild. First, we evaluate how common AVs cope both with unknown malware and well-known evasion techniques. To this end, we develop and use Nuky, a ransomware targeting Windows, and implementing several evasion techniques. 

Second, we design and evaluate a countermeasure, mentioned by \cite{4} and
\cite{11}, that simply consists of reproducing the presence of the artifacts on
computers by instrumenting the Windows API in order to force malware to evade. The purpose of this countermeasure is to limit the infection's spread and not to replace malware detection. We study the impact of this approach on both malware and legitimate software before concluding on its limitations. To evaluate these experiments, we create a small dataset and discuss the problem of collecting evasive samples.

We give an overview of AV abilities and evasion techniques in Section~\ref{sec::abilities} and Section~\ref{sec::evasions}.
We evaluate AV's abilities to detect new evasive malware in Section~\ref{sec::experiments}.
Then, we present the countermeasure and the construction of the dataset used for its evaluation in Section~\ref{sec::dataset} and Section~\ref{sec::counter}.
Finally, before concluding, we discuss the limitations of this study in Section~\ref{sec::limitations}.

\section{\uppercase{State of the art}}
\label{sec::state}
There are multiple definitions of evasive malware in the literature. For instance, \cite{12} define \emph{environment-reactive malware} as:
\begin{displayquote}
  Detection-aware malware carrying multiple payloads to remain invisible to any protection system and to persist for a more extended period. The environment
  aware payloads determine the originality of the running environment. If the environment is
  not real, then the actual malicious payload is not delivered. After detecting
  the presence of a virtual or emulated environment, malware either terminates its
  execution or mimics a benign/unusual behavior.
\end{displayquote}

However, this definition designates malware aiming at detecting virtual or emulated environments only. We complete this definition using part of the proposition made by \cite{6} that calls a malware detecting \textquote{dynamic
  analysis environments, security tools, VMs, etc., as anti-analysis malware}.

We choose to use this definition for the term ``evasive malware'' instead of the
proposed terms ``environment-reactive malware'' or ``anti-analysis malware'',
because we believe this is the term that is the most used by both researchers and malware
writers.

One possibility for malware to detect unwanted environments is to search for
specific artifacts present only in these environments. For instance, \cite{1}
detail an experiment that extracts fingerprints about AV's emulators, and
\cite{16} extract predefined features from a Windows computer. Once the
harvest of artifacts completed, malware can then decide whether they are
running in an AV environment or a standard environment.

Malware then use these artifacts to create evasion tests, as described by
\cite{3}, \cite{5} and \cite{15}. It is also possible to artificially add
artifacts by using implants as detailed by \cite{14}.

To detect evasive malware, we can compare the malware's behavior launched in an analysis environment and on a bare-metal environment. To compare behaviors, \cite{9} search for differences in the system call traces. \cite{8} extract
raw data from the disk of each environment and compare file and registry
operations, among others, to detect a deviation of behavior. Finally, \cite{7} compare the behavior of malware between multiple analysis sandboxes and discover multiple evasion techniques.

It is possible to create an analysis environment that is indistinguishable from
a real one, called a transparent environment, as presented by \cite{17}.
However, \cite{18} gives reservations about the possibilities of creating a
fully transparent environment as for them, \textquote{virtual and native 
hardware are likely to remain dissimilar}.

For this reason, we decide to do the opposite of a transparent environment by
creating an environment containing artificial dissimilarities. To the best of
our knowledge, this idea was already discussed by \cite{4} but never tested on
real malware.

\section{\uppercase{Antivirus abilities}}
\label{sec::abilities}
\noindent AVs are tools that aim at detecting and stopping the execution of new malware samples. They are an aggregate of many features designed to detect the malware's presence by different means, as described by \cite{2}. Using these features incurs overhead that, if too significant, can hinder the use of legitimate programs.  In order to stay competitive, AV editors have to limit this overhead. For this reason, it is difficult for them to apply time demanding techniques.

Most AVs use signature checking by comparing the file against a database of specific strings, for instance. The signature database needs to be updated regularly to detect the latest discovered malware.\footnote{\href{https://support.avast.com/en-us/article/22/}{https://support.avast.com/en-us/article/22/}}\footnote{\href{https://www.kaspersky.co.uk/blog/the-wonders-of-hashing/3629/}{https://www.kaspersky.co.uk/blog/the-wonders-of-hashing/3629/}}
Fewer AVs use features that produce significant overheads, such as running the
malware in an
emulator.\footnote{\href{https://i.blackhat.com/us-18/Thu-August-9/us-18-Bulazel-Windows-Offender-Reverse-Engineering-Windows-Defenders-Antivirus-Emulator.pdf}{https://i.blackhat.com/us-18/Thu-August-9/us-18-Bulazel-Windows-Offender-Reverse-Engineering-Windows-Defenders-Antivirus-Emulator.pdf}}\footnote{\href{https://eugene.kaspersky.com/2012/03/07/emulation-a-headache-to-develop-but-oh-so-worth-it/}{https://eugene.kaspersky.com/2012/03/07/emulation-a-headache-to-develop-but-oh-so-worth-it/}}

During its installation, an AV adds files and modifies the guest OS. For
instance, an AV can load its features by manually allocating memory pages or by
using the \emph{LoadLibrary} API to load a \emph{DLL} file, as described by
\cite{2}. As we detail in the following section, evasive malware can detect these modifications and then adapt their behavior.

Once a new malware is detected, AVs send the sample to their server for further analysis. To analyze malware more deeply, an analyst can use multiple tools such as debugger to break at particular instructions or Virtual Machines (VMs) and emulators to analyze it in a closed environment.

A malware willing to last longer can then try to complicate the analyst's work by using anti-debugger or anti-VM techniques to slow down the analysis.

\section{\uppercase{AV evasions}}
\label{sec::evasions}
In the previous section, we gave details about the environment in which we are interested. We now detail possible ways for evasive malware to detect the presence of such environments. Evasive malware can use many evasion techniques, and \cite{3}, \cite{13} and \cite{5} report a vast majority of them. In this article, we separate these techniques into three main categories: detection of debugging techniques, detection of AV installation and execution artifacts, and detection of VMs. In this section, we briefly describe these categories, and we list the artifacts searched by Nuky for each of them in Table~\ref{tab::summary}.

\paragraph{Detection of debuggers:}
To observe malware in detail, an AV can use a debugger to add breakpoints at a specific instruction. The usage of debuggers implies that a process has to take control of the debugged process. This take of control leaves marks visible from the debugged process, thus can be seen by the malware. For instance, these marks can be the initialization of specific registers, the presence of well-known debuggers' process, or specific behaviors of some Windows API functions.

\paragraph{Detection of AVs' installation and execution artifacts:}
The installation of the AV makes some changes to the operating system. For
Windows, it starts by creating a folder, often in the \emph{Program Files}
folder. This folder contains the executable of the antivirus and the other
resources it requires. During the
installation of an AV, some of the registry keys are
changed or added. They can set up hooks of the Windows
API from the user or the kernel level.

In the same way, the execution of the core process and plugins of the AV leaves artifacts within the operating system, and these alterations can be visible from the user-mode. A simple technique to detect the presence of antivirus is thus to search for standard AV processes names in the list of the running processes.

\paragraph{Detection of VMs, emulators, sandboxes:}
Finally, an AV can execute malware in a sandbox, \emph{i.e.}, a controlled and contained testing environment, where its behavior can be carefully analyzed. These sandboxes intend to reproduce the targeted environment but, indeed, slightly diverge from the real environment. An evasive malware can then search for precise files, directories, process names, or configurations that indicate the presence of a VM or emulator.

\begin{table}[h]
	\vspace{0.2cm}
	\caption{Artifacts searched by Nuky for each category.}
	\vspace{-0.1cm}
	\begin{center}
		\begin{small}
			\begin{tabular}{|C{3.45cm} | C{1.2cm} C{0.4cm} C{0.4cm}|}
				\hline
				Artifacts & Debugger & AV & VM\\
				\hline
				Process Names & $\times$ & $\times$ & $\times$ \\
				GUI Windows Names & $\times$ & & \\
				Debugger registers values & $\times$ & & \\
				Imported functions & $\times$ & & $\times$ \\
				Registries Names \& Values & $\times$ & & \\
				Folder Names & & $\times$ & $\times$ \\
				.DLL Names & & & $\times$ \\
				Usernames & & & $\times$ \\
				MAC addresses & & & $\times$ \\
				\hline
			\end{tabular}
		\end{small}
	\end{center}
	\label{tab::summary}
\end{table}

\section{\uppercase{Experiments}}
\label{sec::experiments}
\subsection{Nuky, a Configurable Malware}
\begin{figure}
	\begin{center}
		\includegraphics[scale=0.6]{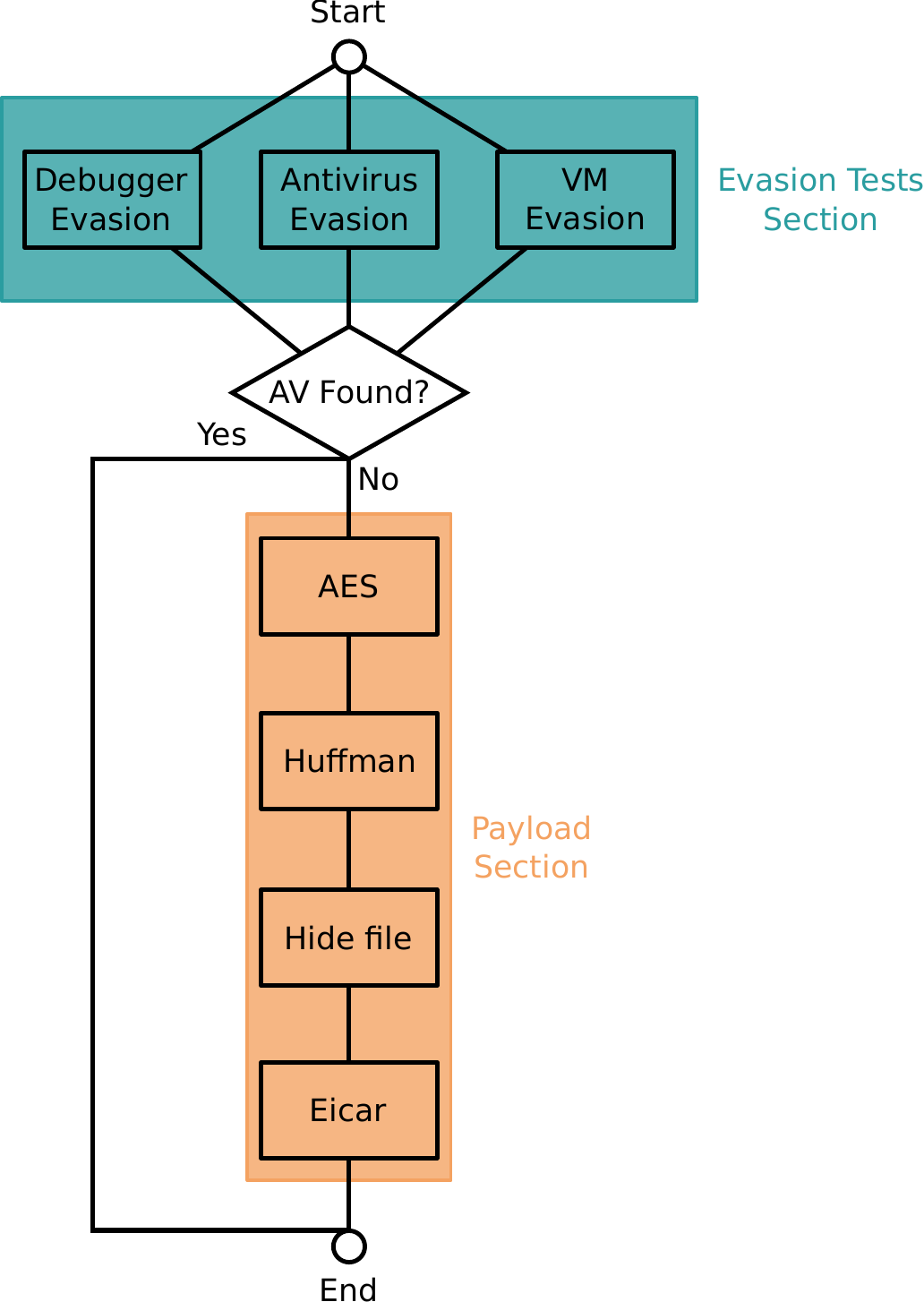}
		\caption{Operation of Nuky.}
		\label{fig::schema-nuky}
	\end{center}
\end{figure}

At this point, we want to measure the malware's ability to evade the most
commonly used AVs. To carry out our experimentations, we develop Nuky, a
configurable malware. Nuky has 4 different payloads and multiple evasion
capabilities divided into 2 sections, as depicted in Figure~\ref{fig::schema-nuky}:

\begin{itemize}
\item The Evasion Tests section is composed of 3 blocks containing numerous
  evasion techniques designed to detect the artifacts listed in Table~\ref{tab::summary}. Nuky is configurable to launch zero or multiple evasion blocks.

\item The Payload section contains 4 payloads detailed in Table~\ref{tab::payloads}. They implement methods to modify
  files in the \emph{Documents} folder on a Windows workstation.
\end{itemize}

\begin{savenotes}
	\begin{table}
		\caption{Nuky's payloads.}
		\vspace{-0.1cm}
		\begin{center}
			\begin{small}
				\begin{tabular}{| c | c |}
					\hline
					Type & Method\\
					\hline
					Encryption & AES in ECB mode\footnote{\href{https://github.com/SergeyBel/AES/}{https://github.com/SergeyBel/AES/}} \\
					Compression & Huffman\footnote{\href{https://github.com/nayuki/Reference-Huffman-coding}{https://github.com/nayuki/Reference-Huffman-coding}}\\
					Stash & Sets hidden attribute to true\\
					Test & Drops Eicar's file on Desktop\footnote{\href{https://www.eicar.org/?page\_id=3950}{https://www.eicar.org/?page\_id=3950}}\\
					\hline
				\end{tabular}
			\end{small}
		\end{center}
		\label{tab::payloads}
	\end{table}
\end{savenotes}

In the following, we use 2 different configurations of Nuky, first to test the detection capability of the available AVs and then to test the evasive capability of malware using the theoretical techniques enumerated in Section~\ref{sec::evasions}.

\paragraph{Set-Up:}
All of our experiments use the same architecture. We prepare Windows images
equipped with some of the most popular AVs, according to a study from
AVTest.\footnote{\href{https://www.av-test.org/en/antivirus/home-windows/}{https://www.av-test.org/en/antivirus/home-windows/}}
These images are all derived from the same Windows 10 base image, populated with
files taken from the Govdocs1
dataset.\footnote{\href{http://downloads.digitalcorpora.org/corpora/files/govdocs1/}{http://downloads.digitalcorpora.org/corpora/files/\\govdocs1/}}
Every AV is set on a high rate detection mode to give more chances for AVs to
trigger alerts. Some AVs propose a specific feature designed to put a specific
folder (specified by the user) under surveillance. For these AVs, we build an
additional Windows image with the AV parametrized to watch the \emph{Documents}
folder. For all our experiments, we update the AVs before disconnecting the
test machine from the network (offline mode), to ensure that the Nuky samples
do not leak to the AVs.

\subsection{AVs' Detection Abilities}
This first experiment aims at finding which AVs can catch Nuky depending on the payload used, and to verify that they are all able to detect the Eicar's file. In this experiment, Nuky does not use any evasion technique and executes each of its 4 payloads. Nuky is a new malware and consequently not present in AVs records. Nevertheless, its first 3 payloads (Compression, Encryption, and Stash) are characteristic of common malware. An AV should, therefore, be able to identify them. Before launching Nuky, we run targeted scans on Nuky's executable file, but no AV succeeds to detect it as malicious. We then execute Nuky once for each payload on every AV with an image restoration step before each run. 

As depicted in Table~\ref{tab::cipher}, only AVs equipped with a specific ransomware detection feature succeed in detecting Nuky's payloads. K7 Computing's product also detects Nuky after some time. Nevertheless, Nuky succeeds in ciphering part of the computer's files before being stopped. We expect K7 Computing to generate false positives as a legitimate image rotation made by the Photos program is detected malicious by this AV. Same observation for Windows Defender blocking a file saved using LibreOffice Draw. However, they still detect Nuky as malicious and are the AVs we want to precisely identify in the second experiment to abort Nuky's execution and avoid detection. Finally, all AVs achieve to detect the test payload with Eicar's file.

The fact that a majority of AVs do not detect the compression, encryption, and stash payloads could be a sign that they still heavily rely on signature analysis.

\begin{table}
	\vspace{0.2cm}
	\caption{Payloads detected by AVs.\\ *Part of the files modified before stopping Nuky.}
	\begin{center}
		\begin{small}
			\begin{tabular}{| c | C{0.3cm} C{0.9cm} C{0.3cm} c|}
				\hline
				Antivirus&AES&Huffman&Hide&Eicar\\
				\hline
				Windows Defender &  &  &  & $\times$\\
				Windows Defender+ & $\times$  & $\times$ & $\times$ & $\times$\\
				Immunet (ClamAV) & & & & $\times$\\
				Kaspersky & & & & $\times$\\
				Avast & & & & $\times$\\
				Avast+& $\times$ & $\times$ & $\times$ & $\times$\\
				AVG & & & & $\times$\\
				Avira & & & & $\times$\\
				K7 Computing & $\times^{*}$ &  $\times^{*}$ &  $\times^{*}$ & $\times$\\
				\hline
			\end{tabular}
		\end{small}
	\end{center}
	\label{tab::cipher}
\end{table}

\subsection{Nuky's Evasion Abilities}
In the second experiment, we enable the three evasion blocks one-by-one, to check whether Nuky using the test payload (Eicar's file) is still detected or not, compared to the first experiment. If Nuky is no longer detected, it means that the evasion block successfully identifies an AV and does not launch its payload. Otherwise, if the evasion block fails, this means that Nuky does not detect the presence of any AV and runs its malicious payload. 

We consider an evasion as successful if at least one of the 2 following conditions is satisfied:
\begin{itemize}
\item An AV detects Nuky in the first experiment, but no longer in the second experiment.
\item Nuky prints the AV's name identified in a console. It can also send the name through an IRC channel.
\end{itemize}

However, as Nuky can be executed in closed environments without a console or
disconnected from the network, we can lose some proofs of evasion.

It may be our case, as only the AV evasion block provides detections for which we can see Nuky's results printed in the console. We do not observe a difference using the debugger evasion block nor the VM evasion block. The reason could be that the AVs we tested do not use the debuggers and VMs targeted by Nuky. It could also be possible that Nuky prints the results on an unreachable console.

Table~\ref{tab::bs} shows the number of artifacts found in each AV by the 2 techniques composing the AV evasion block.

As our evasions heavily rely on string comparisons, our string set has to be exhaustive to detect a maximum of AVs' artifacts. 

Using these evasion methods, we can successfully identify the AVs on which Nuky can launch its ciphering payload without being detected and, thus, increase its life expectancy.

\begin{table}
	\vspace{0.2cm}
	\caption{Number of artifacts detected by Nuky's AV evasion block.}
	\vspace{-0.1cm}
	\begin{center}
		\begin{small}
			\begin{tabular}{|c | C{1.5cm} C{1.5cm}|}
				\hline
				Antivirus & Folder Names & Process Names \\
				\hline
				Windows Defender & 3 & 1\\
				Immunet (ClamAV) & 1 &  1\\
				Kaspersky & 4 & 4\\
				Avast & 2 & 2\\
				AVG &  2 & 2\\
				Avira & 1 & 9\\
				K7 Computing & 2 & 9\\
				\hline
			\end{tabular}
		\end{small}
	\end{center}
	\label{tab::bs}
\end{table}

In the first experiment, we find that AVs have troubles to detect Nuky's payloads unless they propose specific features to detect ransomware. In the second experiment, we show that it is possible to identify precisely AVs to adapt the malware's behavior.

\section{\uppercase{Dataset creation}}
\label{sec::dataset}
\noindent In the previous section, we conclude that the AVs' evasion methods are easy to set up and very efficient against modern AVs. We now detail the way we collected evasive samples in order to build a dataset to test our countermeasure. First, we describe the way we build a filter to keep only evasive samples, and second, how we use it to collect samples.

\subsection{Filtering Using Yara Rules}
To create our filter, we use a program called Yara\footnote{\href{http://virustotal.github.io/yara/}{http://virustotal.github.io/yara/}}, which allows matching samples according to a set of rules. Yara can match samples containing a specific string or loading a specific Windows library or function and, thus, does not work on ciphered or packed malware. However, we believe that the evasive tests could occur before the unpacking or deciphering, thus readable and matchable. 

We create a rule containing five subrules, each composed of several conditions and designed to find malware that respectively try to \emph{detect a debugger, detect a running AV or sandbox, manipulate and iterate over folders, find VM's MAC addresses, and find a debugger window}.

A set of rules designed to match evasive samples already exists\footnote{\href{https://github.com/Yara-Rules/rules/blob/master/antidebug_antivm/antidebug_antivm.yar}{https://github.com/Yara\-Rules/rules/blob/master\\/antidebug\_antivm/antidebug\_antivm.yar}}. However, we create a new one as these rules match too easily and return too much non-evasive malware.

\subsection{Samples Collection}
The difficulty of collecting samples of evasive malware lies in the fact that, by definition, these malware avoid being caught. 
We manually crawl the WEB in public datasets such as hybrid-analysis\footnote{\href{https://www.hybrid-analysis.com/}{https://www.hybrid-analysis.com/}} and public 
repositories like
theZoo\footnote{\href{https://github.com/ytisf/theZoo}{https://github.com/ytisf/theZoo}},
for instance. The data collection covers 3 months between the November 18th,
2019 and the January 21st, 2019.

Over this period, we found 62 samples matching the Yara rule among the few thousand samples crawled.
To avoid false positives and to be sure to keep samples PE files that are more likely to be actual malware, we selected only samples that are not signed and marked as malicious by VirusTotal\footnote{\href{https://www.virustotal.com}{https://www.virustotal.com}}.
Table~\ref{tab::selected} lists all the MD5 of the 18 selected malware, and Table~\ref{tab::matches}, the number of samples matched by each rule.

The fact that some rules do not give any result might be because we crawled the WEB by hand, and therefore, had not numerous samples to match.

The Yara rule we use is very restrictive and matches only a few samples for the duration of the malware harvest. However, we get only a few false positives and now intend to automate the WEB crawling process to continue the acquisition of malware.

\begin{table}
	\vspace{0.2cm}
	\caption{Number of samples matched by each rule.}
	\begin{center}
		\begin{small}
			\begin{tabular}{|c|c|c|}
				\hline
				Rules & All samples & Final samples\\
				\hline
				Debugger & 20 & 16 \\
				AV and sandbox & 0 & 0 \\
				Folder manipulation & 5 & 0 \\
				MAC addresses & 37 & 2 \\
				Find window & 0 & 0 \\
				\hline
			\end{tabular}
		\end{small}
	\end{center}
	\label{tab::matches}
\end{table}

\subsection{Selected Samples Description}
We create groups of samples sharing most of their assembly code by comparing them using Ghidra's diff tool. All samples within a group contain the same code in the text section that seems to carry the evasion part of the malware.

Groups A, B, and C share a significant part of their text section with small additions of code or function's signature modifications.
The other groups are very different and contain their unique code.

We compute the entropy for each sample and find a min-entropy of 6.58 and a max-entropy of 7.11.
Based on the experiment described by \cite{19}, these values could suggest that the samples are packed or encrypted. 
We found in every sample different resources that seem to be PE files that could be unpacked or decrypted after the evasion tests.

All these similarities could be due to the sharing of code on public projects such as Al-Khaser\footnote{\href{https://github.com/LordNoteworthy/al-khaser}{https://github.com/LordNoteworthy/al-khaser}} or smaller ones\footnote{\href{https://github.com/maikel233/X-HOOK-For-CSGO}{https://github.com/maikel233/X-HOOK-For-CSGO}}.
Some packers also propose the addition of an evasion phase before unpacking such as the Trojka Crypter.\footnote{MD5:  c74b6ad8ca7b1dd810e9704c34d3e217 }

\begin{table}
	\caption{Selected samples and countermeasure results.}
	\begin{center}
		\begin{small}
			\begin{tabular}{|C{0.5cm} C{4.7cm} C{0.9cm}|}
				\hline
				Group & MD5 & Results\\ 
				\hline
				\multirow{7}{*}{A} & de3d1414d45e762ca766d88c1384e5f6 & OK\\ 
				& 2d57683027a49d020db648c633aa430a & OK\\
				& d3e89fa1273f61ef4dce4c43148d5488 & OK\\
				& bd5934f9e13ac128b90e5f81916eebd8 & OK\\
				& 512d425d279c1d32b88fe49013812167 & OK\\
				& 2ccc21611d5afc0ab67ccea2cf3bb38b & OK\\
				& 6b43ec6a58836fd41d40e0413eae9b4d & OK\\
				\hline
				\multirow{6}{*}{B} & ee12bbee4c76237f8303c209cc92749d & OK\\ 
				& 5253a537b2558719a4d7b9a1fd7a58cf & OK\\
				& 8fdab468bc10dc98e5dc91fde12080e9 & OK\\
				& e5b2189b8450f5c8fe8af72f9b8b0ad9 & OK\\
				& ee88a6abb2548063f1b551a06105f425 & OK\\
				& 4d5ac38437f8eb4c017bc648bb547fac & OK\\
				\hline
				\multirow{1}{*}{C} & f4d68add66607647bf9cf68cd17ea06a & OK\\
				\hline
				\multirow{1}{*}{D} & 862c2a3f48f981470dcb77f8295a4fcc & CRASH\\
				\hline
				\multirow{1}{*}{E} & e51fc4cdd3a950444f9491e15edc5e22 & NOK\\
				\hline
				\multirow{1}{*}{F} & 812d2536c250cb1e8a972bdac3dbb123 & NOK\\
				\hline
				\multirow{1}{*}{G} & 5c8c9d0c144a35d2e56281d00bb738a4 & CRASH\\
				\hline
			\end{tabular}
		\end{small}
	\end{center}
	\label{tab::selected}
\end{table}

\section{\uppercase{AV evasion countermeasure}}
\label{sec::counter}
\noindent It is possible to thwart such evasion techniques by instrumenting the Windows API to force malware to evade, as discussed by \cite{4}. To do these instrumentations, we used Microsoft Detours, a framework that facilitates the hooking of Windows API functions.\footnote{\href{https://github.com/microsoft/Detours/wiki/OverviewInterception}{https://github.com/microsoft/Detours/wiki/\\OverviewInterception/}}

Microsoft Detours produces a new DLL file loaded at execution time in the malware using the \emph{AppInit} registry\footnote{\href{https://attack.mitre.org/techniques/T1103/}{https://attack.mitre.org/techniques/T1103/}}. This method is efficient against software such as Al-Khaser but needs tests on real malware.
We now describe the Windows API functions that we modify, discuss the countermeasure's efficiency on real malware, and measure the overhead it induces.

\subsection{Windows API Instrumentations}
We implemented 3 types of instrumentations on 8 functions. First, for \emph{IsDebuggerPresent} and \emph{GetCursorPos}, we always return the same fixed value, respectively \emph{True} and a pointer to a cursor structure with the coordinates set to 0.

Second, for \emph{GetModuleHandle}, \emph{RegOpenKeyEx}, \emph{RegQueryValueEx}, \emph{CreateFile}, and \emph{GetFileAttributes}, they access file handles, registry names, and values by providing their names in parameters. We instrument the functions to return fake handles,  values, and system error codes if the name requested is related to an analysis tool.

Finally, \emph{CreateToolhelp32Snapshot} is modified to create fake processes of popular analysis tools if they are not already running on the machine.

\subsection{Countermeasure Efficiency}
\noindent We test our countermeasure by comparing the behavior of the malware we collected on images with and without the countermeasure enabled. We use the same Windows image created for the AVs experiments and create a second image, similar but with the countermeasure installed. We consider that we successfully forced a malware to evade if it behaves differently on the 2 images. All 18 malware are tested by hand using Process Monitor from the \emph{SysInternals} library.\footnote{\href{https://docs.microsoft.com/en-us/sysinternals/downloads/procmon}{https://docs.microsoft.com/en-us/sysinternals/downloads/procmon}}

For 14 malware, we observe on the image without the countermeasure that they launch a subprocess in the background. Besides, these malware start a \emph{WerFault.exe} process.
With the countermeasure enabled, the parent process of these malware immediately stops, which could mean they evade.
The results detailed in Table~\ref{tab::selected} show that 2 malware crash in both environments and 2 others keep the same behavior.

\subsection{Countermeasure Overhead}
We evaluate the overhead induced by the countermeasure on 4 software typically found on Windows computers that are VLC, Notepad, Firefox, and Microsoft Paint. We also measure Al-Khaser, a program performing a set of evasion techniques and generating a report on their results. We removed from Al-Khaser all the techniques relying on time observation.

We run each software 100 times, interacts with them, and computes the mean of all the execution times with and without the countermeasure enabled, as shown in Table~\ref{tab::overhead}.
The only significant overhead that we observe is on Al-Khaser with +14.62\%, as this software heavily calls our instrumented functions, which we consider the worst-case scenario. 

\begin{table}
	\vspace{0.2cm}
	\caption{Executions times (in seconds) with and without the countermeasure enabled. The first line of each row is with the countermeasure enabled, and the second line without.}
	\begin{center}
		\begin{small}
			\begin{tabular}{|c|c|c|}
				\hline
				Software & Mean & Standard Deviation\\
				\hline
				\multirow{2}{*}{Al-Khaser} & 15.24 & 1.38\\
				& 13.29 & 1.36\\
				\hline
				\multirow{2}{*}{VLC}& 0.8772 & 0.0609\\
 				& 0.8593 & 0.0677\\
				\hline
				\multirow{2}{*}{Notepad}& 0.0615 & 0.0033\\
				& 0.0607 & 0.0028\\
				\hline
				\multirow{2}{*}{Firefox}& 0.8005 & 0.0270\\
				& 0.8026 & 0.0259\\
				\hline
				\multirow{2}{*}{MS-Paint}& 0.0581 & 0.0096\\
				& 0.0583 & 0.0106\\
				\hline
			\end{tabular}
		\end{small}
	\end{center}
	\label{tab::overhead}
\end{table}

\section{\uppercase{Limitations and Future Work}}
\label{sec::limitations}
\subsection{Nuky}
When testing AVs against evasion techniques, we use a minimal number of basic evasion techniques, well known to attackers, and easily reproducible. Almost all the techniques we test are based on string comparisons and are not representative of all the possible ways to detect an AV. For now, Nuky represents the malware that any attacker could write by looking at tutorials, and not the elaborated malware created by big groups or states. We intend to extend Nuky's functionalities to make it harder to detect.

We did not achieve to test the evasion techniques targeting debuggers and VM's with our simple experiment. We can then only conclude on the efficiency of the category of evasions targeting AVs, which contains for now 2 techniques. However, we believe that it is possible to test the other categories of evasion with more elaborated black-box testing.

\subsection{Countermeasure}
We chose to instrument the Windows API functions to block the same basic techniques used by Nuky and not more elaborated ones. As for a signature database, this countermeasure needs continuous updates to be able to thwart new evasive techniques. Of course, with this method, we can only block malware evading using the Windows API and not those using different means.

To avoid testing the countermeasure on malware it was designed for, we used a different set of strings to construct the dataset and to implement the countermeasure. However, this bias is still present and is removable by using different techniques to collect malware and to evaluate the countermeasure.

We also need to measure the side-effects of this countermeasure on malware, legitimate software, and the operating system.

\subsection{Dataset}
There is no public repository properly providing malware with the details of the evasion techniques performed.
For this reason, we use a small dataset we created and analyzed by hand, but on which we cannot generalize our conclusion for now.

We aim at automatically crawling and flagging samples to create a larger dataset and observe the evolution of such evasive malware through time by letting this experiment run in the long term.

Finally, after obtaining a fully formed dataset of evasive malware, we intend to study how they evade. Most of the tested malware just stop their execution, but others could choose to behave differently while still executing.

\section{\uppercase{Conclusion}}
\label{sec::conclusion}
\noindent We found that AVs are still vulnerable to unknown malware using basic encryption, compression, and attribute modification. Only 3 out of the 9 AVs we tested caught Nuky, and we suspect them to generate a lot of false positives. Other AVs may also still rely on simple signature detection. 

Moreover, when an AV succeeds in catching our malware, we prove that basic evasive techniques enable us to identify them precisely and change the malware's behavior to avoid detection. We showed that Nuky could escape AVs by detecting AV artifacts but need more tests for the debugger and VM artifacts.

A few samples of real malware implementing evasion techniques are collected using Yara rules. We found that a lot of these samples share their evasive code that we retrieved in public code repositories.

Finally, we implemented a countermeasure that seems to be able to thwart evasive malware by instrumenting the Windows API. In the end, 14 out of the 18 malware tested showed a different behavior with and without the countermeasure enabled. The highest overhead measured is on Al-Khaser with an addition of 14.62\% to the execution time.

The code for Nuky's evasion part and the Yara rule is available on request.
\bibliographystyle{apalike}
{\small
\bibliography{main}}

\begin{thebibliography}{}

\bibitem[Afianian et~al., 2020]{5}
Afianian, A., Niksefat, S., Sadeghiyan, B., and Baptiste, D. (2020).
\newblock Malware dynamic analysis evasion techniques: A survey.
\newblock {\em {CSUR} Computing Surveys - ACM}, 52(6).

\bibitem[Blackthorne et~al., 2016]{1}
Blackthorne, J., Bulazel, A., Fasano, A., Biernat, P., and Yener, B. (2016).
\newblock Avleak: Fingerprinting antivirus emulators through black-box testing.
\newblock In {\em {WOOT} Workshop on Offensive Technologies}, number~10, pages
  91--–105, Austin, TX, USA. {USENIX} Association.

\bibitem[Bulazel and Yener, 2017]{3}
Bulazel, A. and Yener, B. (2017).
\newblock A survey on automated dynamic malware analysis evasion and
  counter-evasion: Pc, mobile, and web.
\newblock In {\em {ROOTS} Reversing and Offensive-Oriented Trends Symposium},
  number~1, pages 1--21, Vienna, Austria. {ACM}.

\bibitem[Chen et~al., 2016]{11}
Chen, P., Huygens, C., Desmet, L., and Joosen, W. (2016).
\newblock Advanced or not? {A} comparative study of the use of anti-debugging
  and anti-vm techniques in generic and targeted malware.
\newblock In {\em {IFIP SEC} International Information Security and Privacy
  Conference}, number~31, pages 323--336, Ghent, Belgium. Springer.

\bibitem[Chen et~al., 2008]{4}
Chen, X., Andersen, J., Mao, Z.~M., Bailey, M., and Nazario, J. (2008).
\newblock Towards an understanding of anti-virtualization and anti-debugging
  behavior in modern malware.
\newblock In {\em {DSN} Dependable Systems and Networks}, number~38, pages
  177--186, Anchorage, Alaska, USA. {IEEE} Computer Society.

\bibitem[Dinaburg et~al., 2008]{17}
Dinaburg, A., Royal, P., Sharif, M.~I., and Lee, W. (2008).
\newblock Ether: malware analysis via hardware virtualization extensions.
\newblock In {\em {CCS} Conference on Computer and Communications Security},
  number~15, pages 51--62, Alexandria, Virginia, USA. {ACM}.

\bibitem[Garfinkel et~al., 2007]{18}
Garfinkel, T., Adams, K., Warfield, A., and Franklin, J. (2007).
\newblock Compatibility is not transparency: {VMM} detection myths and
  realities.
\newblock In {\em {HotOS} Hot Topics in Operating Systems}, number~11, pages
  30--36, San Diego, California, USA. {USENIX} Association.

\bibitem[Kirat and Vigna, 2015]{9}
Kirat, D. and Vigna, G. (2015).
\newblock Malgene: Automatic extraction of malware analysis evasion signature.
\newblock In {\em {SIGSAC} Conference on Computer and Communications Security},
  number~22, pages 769--780, Denver, Colorado, USA. {ACM}.

\bibitem[Kirat et~al., 2014]{8}
Kirat, D., Vigna, G., and Kruegel, C. (2014).
\newblock Barecloud: Bare-metal analysis-based evasive malware detection.
\newblock In {\em {USENIX} Security Symposium}, number~23, pages 287--301, San
  Diego, California, USA. {USENIX} Association.

\bibitem[Koret and Bachaalany, 2015]{2}
Koret, J. and Bachaalany, E. (2015).
\newblock {\em The Antivirus Hacker’s Handbook}.
\newblock Number~1. Wiley Publishing.

\bibitem[Lindorfer et~al., 2011]{7}
Lindorfer, M., Kolbitsch, C., and Comparetti, P.~M. (2011).
\newblock Detecting environment-sensitive malware.
\newblock In {\em {RAID} Recent Advances in Intrusion Detection}, number~14,
  pages 338--257, Menlo Park, California, USA. Springer.

\bibitem[Lita et~al., 2018]{13}
Lita, C., Cosovan, D., and Gavrilut, D. (2018).
\newblock Anti-emulation trends in modern packers: a survey on the evolution of
  anti-emulation techniques in {UPA} packers.
\newblock {\em Computer Virology and Hacking Techniques}, 12(2).

\bibitem[Lyda and Hamrock, 2007]{19}
Lyda, R. and Hamrock, J. (2007).
\newblock Using entropy analysis to find encrypted and packed malware.
\newblock {\em {SP} Security {\&} Privacy - IEEE}, 5(2).

\bibitem[Miramirkhani et~al., 2017]{15}
Miramirkhani, N., Appini, M.~P., Nikiforakis, N., and Polychronakis, M. (2017).
\newblock Spotless sandboxes: Evading malware analysis systems using
  wear-and-tear artifacts.
\newblock In {\em {SP} Symposium on Security and Privacy}, number~38, pages
  1009--1024, San Jose, California, USA. {IEEE} Computer Society.

\bibitem[Naval et~al., 2015]{12}
Naval, S., Laxmi, V., Gaur, M.~S., Raja, S., Rajarajan, M., and Conti, M.
  (2015).
\newblock Environment--reactive malware behavior: Detection and categorization.
\newblock In {\em {DPM/QASA/SETOP} Data Privacy Management, Autonomous
  Spontaneous Security, and Security Assurance}, number~3, pages 167--182,
  Wroclaw, Poland. Springer.

\bibitem[Tan and Yap, 2016]{6}
Tan, J. W.~J. and Yap, R. H.~C. (2016).
\newblock Detecting malware through anti-analysis signals - {A} preliminary
  study.
\newblock In {\em {CANS} Cryptology and Network Security}, number~15, pages
  542--551, Milan, Italy. Springer.

\bibitem[Tanabe et~al., 2018]{14}
Tanabe, R., Ueno, W., Ishii, K., Yoshioka, K., Matsumoto, T., Kasama, T.,
  Inoue, D., and Rossow, C. (2018).
\newblock Evasive malware via identifier implanting.
\newblock In {\em {DIMVA} Detection of Intrusions and Malware, and
  Vulnerability Assessment}, number~15, pages 162--184, Saclay, France.
  Springer.

\bibitem[Yokoyama et~al., 2016]{16}
Yokoyama, A., Ishii, K., Tanabe, R., Papa, Y., Yoshioka, K., Matsumoto, T.,
  Kasama, T., Inoue, D., Brengel, M., Backes, M., and Rossow, C. (2016).
\newblock Sandprint: Fingerprinting malware sandboxes to provide intelligence
  for sandbox evasion.
\newblock In {\em {RAID} Research in Attacks, Intrusions, and Defenses},
  number~19, pages 165--187, Evry, France. Springer.

\end{thebibliography}

\end{document}